# Initial Observations of the First BlueBird Spacecraft and a Model of Their Brightness

Richard E. Cole[*1], Anthony Mallama[1], Scott Harrington and Jay Respler[1]

[1] IAU - Centre for the Protection of Dark and Quiet
Skies from Satellite Constellation Interference

* Correspondence: r.e.cole@outlook.com

Date: 2025-05-06

Abstract

Based on a large set of visual observations, the mean apparent magnitude of
BlueBird satellites is 3.44, while the mean of magnitudes adjusted to a uniform
distance of 1000 km is 3.84. Near zenith the spacecraft can be as bright as
magnitude 0.5. While these spacecraft are bright enough to impact astronomical
observations, they can for periods be fainter than the BlueWalker 3 prototype
satellite. A model for their brightness shows that design changes since the
BlueWalker 3 mission can explain the behavior of BlueBird.

## 1    Introduction

Bright spacecraft are an issue for astronomers
because they interfere with observation of the night
sky (Barentine et al. 2023 and Mallama and Young
2021). The large satellites launched by AST
SpaceMobile (AST) are a special concern because
they can be very bright. The luminosity of their
BlueWalker 3 prototype satellite was characterized
by Mallama et al. (2023 and 2024). Now, AST has
launched the initial five spacecraft of their follow-on
BlueBird constellation. This paper reports on their
magnitudes.

Section 2 describes how brightness was measured
and Section 3 characterizes the resulting
magnitudes. Sections 4 and 5 present a physical
model of BlueBird satellites that fits the observations
and explains characteristics of the spacecraft
brightness. Section 6 compares the BlueBird
spacecraft with other satellite constellations and
discusses their impact on astronomy. Section 7
presents our conclusions.

## 2    Observations

Visual magnitudes were determined by comparing
the satellites to nearby reference stars. Most of the
measurements were made using binoculars, with
the remainder naked-eye. The angular proximity
between satellites and stellar objects accounts for
variations in sky transparency and sky brightness.
Mallama (2022a) describes this method in more
detail.

BlueBird satellites unfold from relatively compact objects to large flat-panel antennas sometime after they reach orbit. The magnitude statistics discussed in the following section are derived from 300 visual observations of spacecraft that had unfolded.

## 3 Brightness characterization

An important parameter for assessing how seriously satellites interfere with optical astronomy is their *mean apparent magnitude*. That value for the BlueBird satellites is 3.44 with a standard deviation (SD) of 1.55 and a standard deviation of the mean (SDM) of 0.09. Figure 1 shows that the distribution of BlueBird magnitudes is somewhat fainter that of BlueWalker 3.

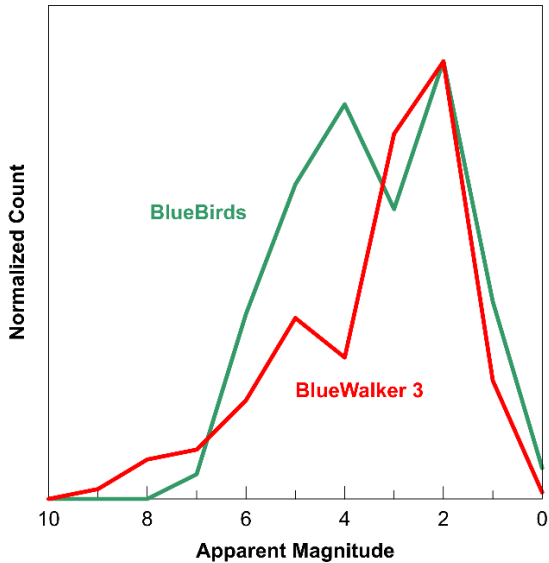

*Figure 1: Distribution of apparent magnitudes.*

Another useful parameter is *the mean of magnitudes adjusted to a uniform distance of 1,000 km*. That allows for comparison between different models of spacecraft which may be flown at different altitudes. The mean, SD and SDM for 1000-km magnitudes for the BlueBirds are 3.84, 1.37 and 0.08. Figure 2 compares the distributions of BlueBird and BlueWalker 3 magnitudes after adjustment to 1,000 km.

*Beta* is the angle measured between the plane of a satellite orbit and the direction to the Sun. Brightness may depend on the beta angle for various reasons. For example, BlueWalker 3 was tilted towards the Sun when the beta angle was high. Figure 3 shows that luminosity is a stronger function of cosine beta for BlueWalker 3 than for the BlueBirds.

The *phase angle* is measured at a satellite between directions to the Sun and to the observer. Many spacecraft are brighter at small phase angles when they are nearly fully lit by the Sun than at large angles when they are mostly backlit. Figure 4 shows that the luminosity of BlueBird satellites is a stronger function of phase angle than that of BlueWalker 3.

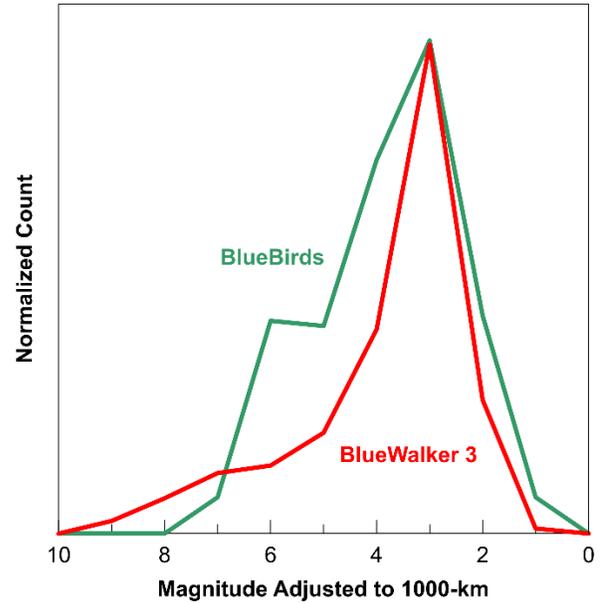

*Figure 2: Distribution of magnitudes adjusted to a uniform distance of 1000 km.*

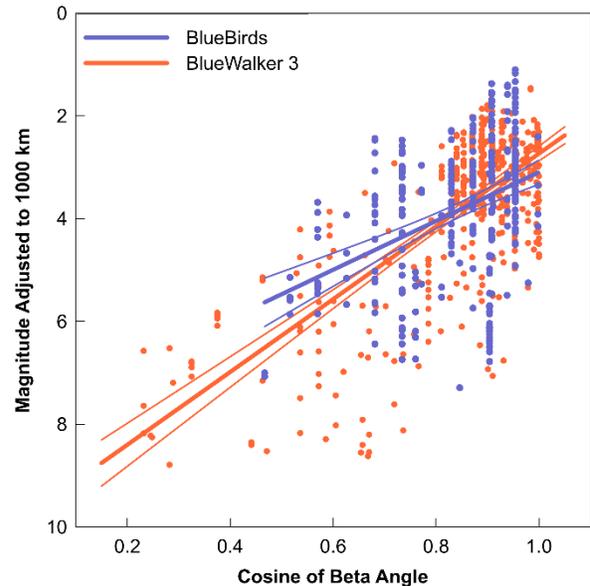

*Figure 3: Brightness as a function of the beta angle.*

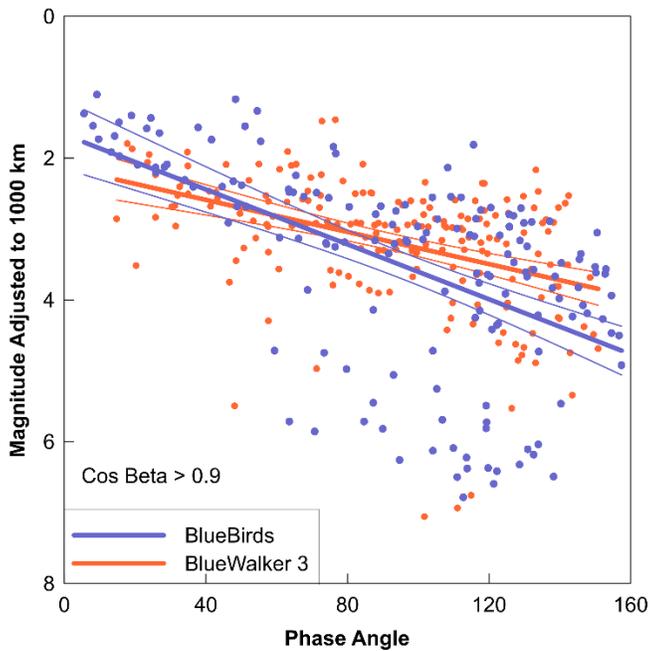

Figure 4: Brightness as a function of phase angle. Observations obtained at large values of beta angle are omitted in order to reduce the influence of the strong beta angle function.

## 4 A brightness model of the BlueBird spacecraft

### 4.1 The BlueWalker 3 model

In the brightness modelling carried out for the BlueWalker 3 spacecraft (Mallama et al 2023), the structure of the spacecraft (Figure 5) could be represented by a simple flat plane diffusely reflecting sunlight.

A reasonable level of accuracy was achieved in the BlueWalker 3 brightness model as shown in Figure 6, taken from that work.

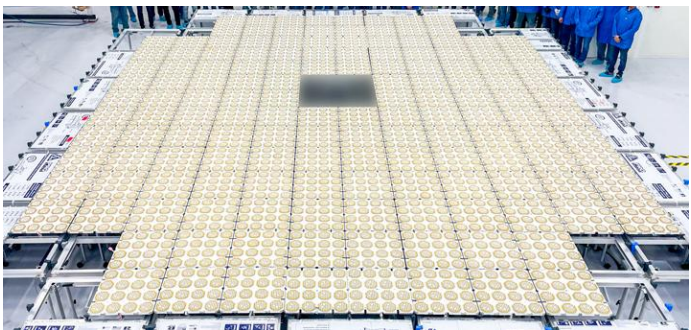

Figure 5: The BlueWalker 3 spacecraft in testing before launch (image from AST)

### 4.2 BlueBird program

Following the BlueWalker 3 launch in 2022, a number of organizations voiced complaints about its brightness in the sky (e.g. Nandakumar, Eggl, Tregloan-Reed et al. 2023).

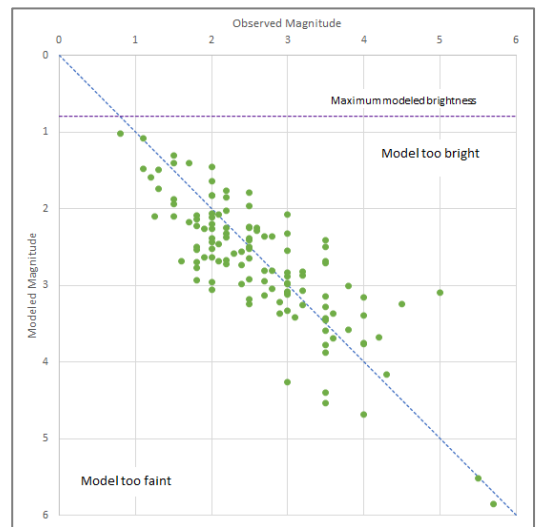

Figure 6: Comparison of the brightness predictions of the BlueWalker 3 brightness model and observations made in the period between the AST announcement of full deployment and 2002 December 7.

Therefore, there was a possibility that AST would make changes in the BlueBird design to reduce its brightness, compared to BlueWalker 3. A review was made of images of the BlueBird production process published on the AST website, in AST financial reports and on Facebook. The most relevant extracted images are shown in Figure 7 and Figure 8. Note that these images are not high resolution as they were clipped from the available sources. They both contain areas that have been blurred out by AST, presumably to conceal proprietary information.

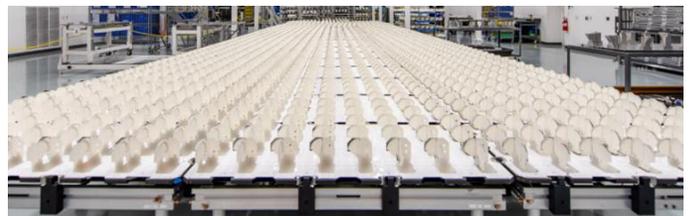

Figure 7: Image taken in a large clean room showing what appears to be a number of BlueBird antenna panels, each termed a micron by AST (image from AST)

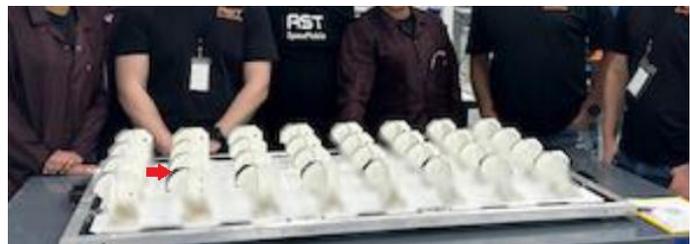

Figure 8: A single micron antenna panel. The dark back-surface material referred to in the text is indicated (image from AST)

The antenna panels or microns have 'standoff structures' that were not present on the BlueWalker 3 panel (Figure 5). Each micron has 32 structures mounted, each of which appears to comprise two 'mushroom slices' mounted at right-angles on a bracket. This structure has the same general coloration in the images as the micron. AST have informed us (AST, 2024b) that each standoff structure carries an antenna element and that these are deployed after launch. AST stated that they do not move again after deployment.

A statement by AST to the US Federal Communications Commission (AST, 2024a) indicate that "deployable antenna elements" are intended to be part of brightness mitigation steps for BlueBird. Assuming the stand-off structures in the images are the stated deployable elements, they would potentially have an effect on the brightness of BlueBird by shadowing sunlight incident on the large Earth-facing antenna panel. The work reported here models the impact of these stand-off structures on the brightness of BlueBird and compares the predictions of the model to the observations.

### 4.3 Attitude control of the spacecraft

AST kindly provided the authors with the following information (AST, 2024b) on the attitude control of the BlueBird spacecraft during the major part of the observation period discussed here.

1) The satellites' pointing direction was tracking the orbit beta angle (zenith face rolled toward the sun) at beta angles up to 20°. Above this beta angle the roll angle was maintained at 20°.

2) The long axis of the spacecraft remained aligned with the orbit velocity-vector. Therefore, the rotation towards the Sun is around the velocity-vector.

The effect of this roll maneuver on the illumination of the nadir panel during observations is moderated by the fact that the spacecraft is only visible from the ground when it is crossing the terminator, 90° from the Sun vector. At this time and if the beta angle is small or zero, the spacecraft velocity-vector is close to the Sun vector and any roll of the spacecraft around the velocity-vector has a small effect of the Sun angle on the nadir panel, Figure 9. As the beta angle becomes greater, the effect of roll on the sunlight angle of incidence increases.

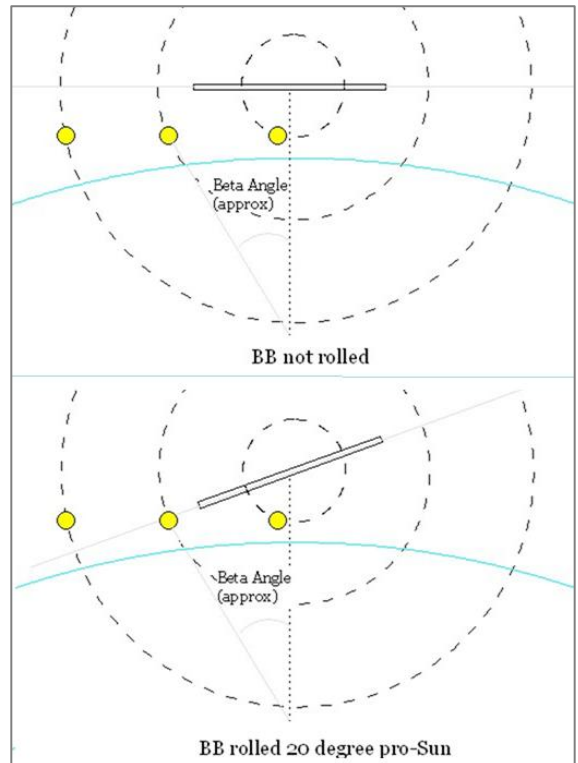

Figure 9: The effect of BlueBird spacecraft roll on the illumination of the nadir panel, when the spacecraft is close to the terminator. The velocity-vector is into the page. The position of the Sun is shown for a number of orbit beta angles.

### 4.4 Ray-tracing

A simple ray-tracing model was developed to understand the effect of the stand-offs on the optical reflection from the BlueBird Earth-facing antenna panel.

The AST animations of the BlueBird spacecraft on-orbit show the folds between the long sections in the antenna panel being held parallel to the velocity-vector of the orbit, consistent with what AST has stated to the authors (Figure 10).

On that basis, the two elements of each standoff-structure are kept parallel to and at right angles to the velocity-vector, respectively. The direction of the Sun illumination on the standoffs will vary from being close to parallel to the velocity-vector when the orbit beta angle is zero and up to about 70° to the velocity-vector when the beta angle is large.

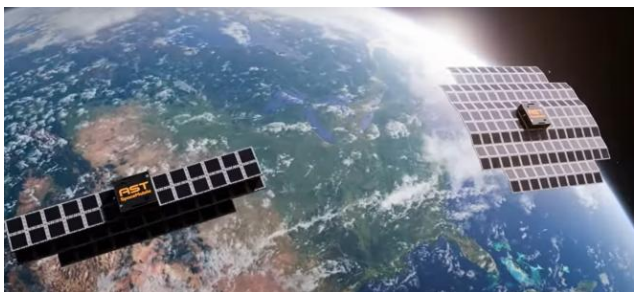

Figure 10: A frame taken from a BlueBird deployment animation indicating its orientation during deployment (image from AST)

Because the spacecraft is near the terminator (the line between day and night on the ground) when it is visible on the ground, the azimuth angle of the Sun from the velocity-vector is very close to the orbit beta angle, but not identical (Figure 11). This relative azimuth difference is termed the *beta-prime angle* in the rest of this paper. The beta-prime angle is computed for each observation.

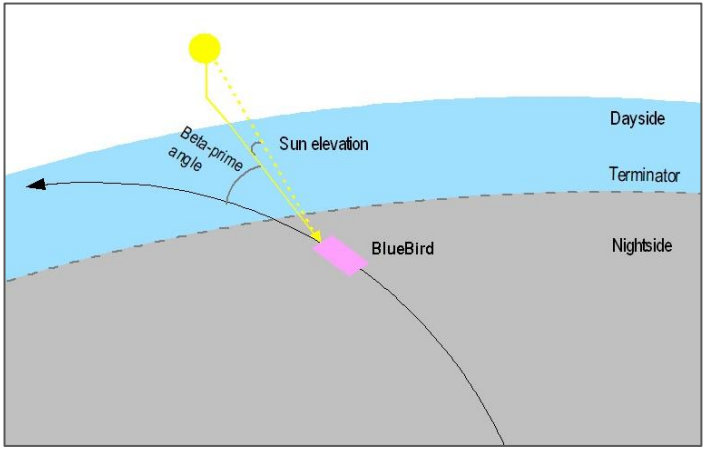

*Figure 11: Schematic of a BlueBird spacecraft crossing the terminator. The beta-prime angle and sun elevation at the spacecraft are marked.*

The Sunlight elevation angle at the spacecraft is in the range 0 to -22° - the spacecraft enters eclipse when this angle reaches -22°. The actual angle of illumination of the Sun on the nadir panel due to the roll angle can be calculated.

Depending on the values of Sun illumination angle on the nadir panel and beta angle, the length and rotation angle of the shadows cast on the antenna panel vary widely. If the shadows of the stand-off structures completely cover the antenna panel, little sunlight will directly reach the antenna panel. Instead, sunlight will reflect from the standoff structures in a direction largely parallel to the antenna panel surface. The simulated illuminated fraction of the antenna panel derived from the ray-tracing is plotted in Figure 12 for several values of beta-prime angle between 0 and 45°, against Sun elevation at the spacecraft. When beta-prime angle is zero (sunlight arriving from directly ahead or behind the spacecraft) then the shadowing is a minimum with the strips between the stand-off structures unshadowed. For values of beta-prime angle greater than 35° the panel is almost completely shadowed by the stand-offs.

The model does not include effects on the edge of the antenna face where the shadowing will be less effective or any areas of the nadir face without the stand-offs, e.g. the control unit at the center of the nadir panel. These effects will mean that the nadir face will be brighter than indicated in Figure 12.

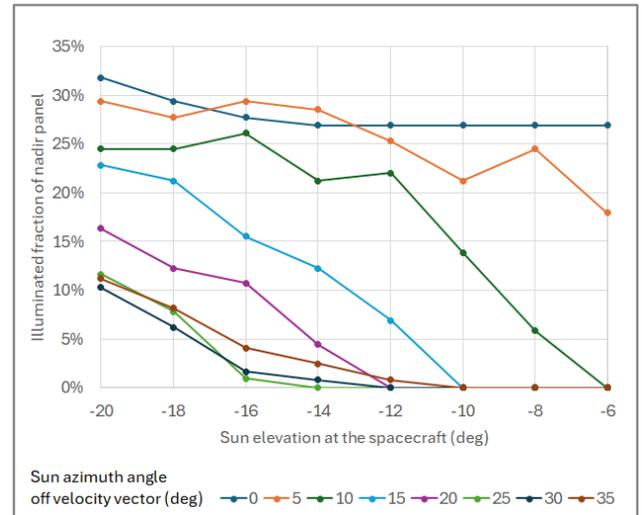

*Figure 12: Fraction of the nadir panel illuminated by the Sun as calculated from the ray tracing model. Sun angles off the velocity-vector from 0 to 35° shown individually.*

The faces of the deployed standoffs will be illuminated by the Sun, the amount of illumination on these faces varying with the Sun azimuth and elevation angles on the nadir panel and the beta-prime angle. The stand-offs will shadow each other, such that when the Sun elevation angle is zero or above the plane of the antenna panel only the stand-offs at the outside edge of the nadir-face will be illuminated. When the spacecraft is rolled, this brightness element is important at all observed elevations.

### 4.5 BlueBird Brightness model

The same brightness modeling methodology was employed as was used on observations of Starlink Visorsats (Cole 2021). In the case considered here, a lot more information is available about the design and orientation of the spacecraft than was for Starlink Visorsats at that time. The purpose of the model is to characterize the brightness of BlueBird so that the effect of various aspects of the design can be individually understood.

The model was developed using the observation set and particularly the following parameters for each observation:

1. The location of the observer (in particular, latitude)
2. The angle of the sun below the horizon at the observer and its azimuth
3. The azimuth and elevation of the spacecraft in the sky
4. The ground track of the spacecraft (which determines the velocity-vector)
5. The orbit beta angle at the time of observation, which determines the roll-angle of the spacecraft

The beta-prime angle used in calculating the shadowing by the deployable antenna elements is individually determined for each observation by the parameters listed above.

The components of the brightness model are:

1. Sunlight scattered from the nadir panel, after allowing for the shadowing by the stand-offs.
2. Sunlight directly scattered by the standoffs faces directed in each of the Ram, Port, Wake and Starboard directions.
3. Earth-light scattered from the nadir panel.

Earth-light is the strong reflected sunlight from the sunlit side of the Earth. It illuminates any object close to the Earth.

These components are shown in Figure 13.

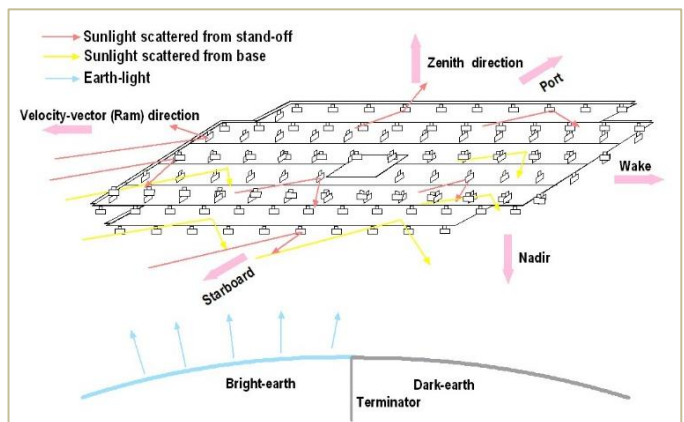

*Figure 13: The elements of the brightness model shown against a rough schematic of the Earth face of the BlueBird spacecraft (not to scale and a small fraction of the actual number of stand-offs is shown).*

## Component 1 - Sunlight scattered from the nadir panel

The ray tracing results described above were used as a model of the illumination of the nadir panel that could be scaled in a fit to the observations.

Simple Lambertian scattering is assumed for the flux from the unshadowed part of the nadir panel.

## Component 2 - Sunlight scattered by the standoffs

The light intercepted by the standoffs is scattered centered on the plane of the antenna panel, at right angles to light scattering from the nadir panel.

For the purposes of calculating their contribution to the brightness of the spacecraft, the vertical faces of the standoffs can be treated as four panels at right-angles. Each of the four sides of this collective is treated separately, calculating the angle of incidence and angle of reflection for each side. Allowance is made for self-shadowing (Figure 14a) which reduces this component as the Sun elevation angle increases until only the edge is illuminated.

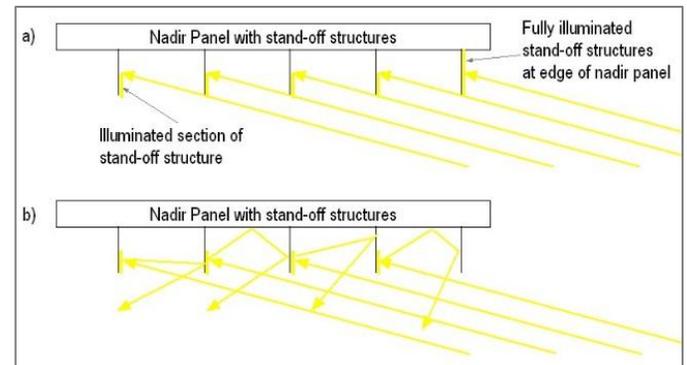

*Figure 14: a) Self-shadowing of stand-off structures that reduces their contribution to the object brightness
b) re-scattering via the stand-off structures and the nadir panel.*

Further, there is the possibility of re-scattering some of this reflected sunlight from the anti-sun sides of the stand-offs (Figure 14b).

This re-scattered sunlight would also be a factor in illuminating sections of the nadir panel that are shadowed from direct sunlight by the stand-offs.

## Component 3 – Earth-light

The BlueBird nadir panel is large and reflective so will reflect earth-light when it is present. A reflected earth-light component may be apparent when spacecraft is within a few degrees of the terminator.

The residuals from fitting the observation dataset were used to derive an earth-light component, discussed below.

## 5  Fitting to the observation data

### 5.1  Model Components

The model is based on five surfaces diffusely reflecting sunlight, so the reflected flux can be calculated using the cosines of the angles of incidence and reflection. The geographical sub-satellite positions and ranges for each observation are calculated from the recorded data for each observation, satellite elevation and azimuth, Sun elevation and azimuth, observer position plus the satellite height derived from current orbital elements.

The surfaces are aligned to the velocity-vector as shown in Figure 13. The velocity-vector is calculated from a model of the orbit, the position of the sub-satellite point and whether the spacecraft is north or southbound. The roll angle of the spacecraft is derived from the beta angle, calculated from the orbital elements and from the AST attitude control rule stated in section 4.3.

List of fitted parameters with best fit values:

| | | |
|---|---|---|
| $A$ | reflectivity x area of nadir panel | 1 |
| $RT_f$ | normalization of unshaded fraction of nadir panel from ray-tracing | 2.1 |
| $SPR$ | reflectivity x area of 'side panels' relative to base (stand-offs) | 0.75 |
| $USH$ | Unshaded fraction of nadir panel | 0.18 |
| $SP_{OP}$ | Relative brightnTHeess of 'side panels' in opposite direction | 0.15 |
| $M_{off}$ | fitted single magnitude offset for all points | -1.8 |
| $SS_c$ | self-shading factor for edge of panel | 0.05 |
| $EL_c$ | Earth-light – normalisation | 1.9 |
| $EL_a$ | Earth-light - Sun elevation for start of contribution | 15° |
| $EL_p$ | Earth-light - power of sine angle | 2.5 |

List of other terms:

| | |
|---|---|
| $EL$ | earth-light contribution |
| $M_{range}$ | Magnitude correction to the range at the observation |
| $RT$ | raytraced unshaded fraction of nadir panel as function of sun elevation and azimuth on panel |
| $SE$ | sun elevation angle on the nadir panel, after roll angle |
| $SS$ | self-shading of stand-offs |
| $i_n$ | solar angle of incidence for surface $n$, $n$ =1-5 |
| $r_n$ | angle of reflection to the observer from surface $n$, $n$ =1-5 |

Elements of the model:

| Element | Equation |
|---|---|
| Nadir panel | $C1 = A * RT * (RT_f + USH) * \cos(i_1) * \cos(r_1)$ |
| Side panels (4) | $C2 = SS * A * SPR *$ $\left( \sum_{n=2}^{n=5} \cos(i_n) * \cos(r_n) + SP_{OP} * \sum_{n=2}^{n=5} \cos(180 - i_n) * \cos(r_n) \right)$ $Cosine\ values\ set\ to\ 0\ if\ angle > 90°$ |
| Earth-light | $C3 = EL_c * \sin^{EL_p}(SE + EL_a)$ |
| SS | $SE < 0, SS = Tan(-SE)/Tan(32°) + SS_c$ $SE > 0, SS = SS_c$ |

The final model for the magnitude is thus:

$$M_{v=} = 2.5 * log_{10}(C1 + C2 + C3) + M_{off} + M_{range}$$

### 5.2  Fitting Process

The fitting process involved iterations to the model based on examination of residuals plotted against a series of parameters, particularly:

- Sun elevation at the spacecraft
- Sun elevation on the nadir panel
- Beta-prime angle
- Phase angle
- Observer view angle to the nadir panel

The parameters adjusted are stated in the previous section.

The objective of the fitting process was to reduce the root-mean-square error to ~0.5ᵐ, the approximate error of visual magnitude measurements. The measurement error can be larger at bright magnitudes as there are few stars in sky to act as comparison objects in visual observations.

### 5.3 Results from the Fitting Process

The dataset can be divided into two parts, observations with beta-prime angle<25° and those with beta-prime angle>25°. The modeled brightness of the former dataset part is dominated by the nadir panel as that is largely unshaded and the roll-angle is ≤20°. The brightness of the latter dataset part is dominated by the side-panels (the stand-offs) as shading of the nadir panel is nearly complete and the roll-angle is 20° - therefore the Sun illumination of the nadir panel is low and the side panels (the stand-offs) are rotated to be more visible to the observer. The key parameter in the model is SPR, the ratio of reflecting power of a side-panel to the nadir panel. Because the stand-offs can reflect away virtually all the sun-light incident on the nadir panel, when the beta-prime angle is large, this parameter must be approaching unity.

This was tested by fitting the beta-prime>25° dataset (138 observations) with values of $SPR$ between 0 and 1, applying a chi-squared test comparing the test statistic with the appropriate critical value for P=0.05. The standard-deviation of the observed brightnesses was taken to be 0.5ᵐ. The results are shown in Figure 15, indicating the value of $SPR$ is >0.45. A best fit value of 0.75 is used in later plots.

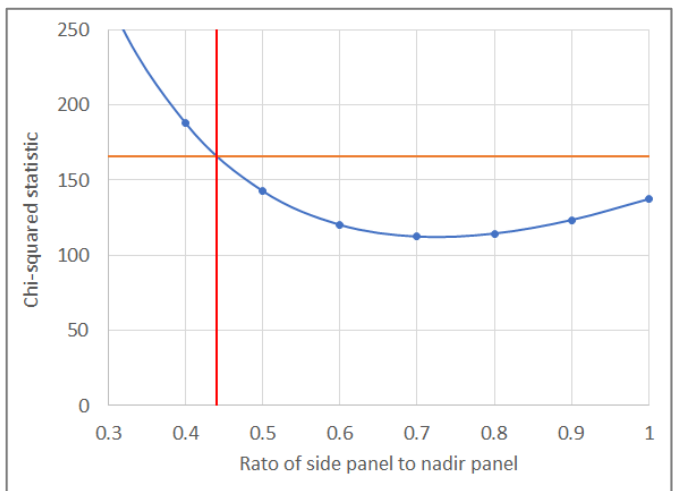

*Figure 15: Chi-squared test result for the brightness of the side-panels relative to the nadir panel*

There must be some always unshaded part of the nadir panel, at the edges and on the Controlsat at the center of the nadir panel. The error distribution with parameter ($USH$) across all observations was asymmetric and improved with $USH$ value of 0.18. The improvement in fit from this value is not statistically significant but it is used here.

Earth-light - For the data where the spacecraft was close to the terminator or over the sunlit side of the Earth, an earth-light component was added with parameters to fit the data.

Reflected earth-light has been previously seen in observations of Gen2 and Direct-to-Cell (DTC) Starlink spacecraft (Mallama et al 2025). It was unsurprising this would also be detected in BlueBird observations.

Building a physical model of the earth-light itself is beyond the scope of this paper, Fankhauser et al (2023) has a detailed treatment of earth-light in their model of the Starlink spacecraft. Here, we have used a simple model based on the Sun elevation at the spacecraft and the angle where earth-light becomes apparent in observations.

Fitted shadowing by the stand-offs – A single parameter ($RT_f$) was included in the model, scaling the ray-tracing results to allow for the areas of the nadir panel that were not shadowed as efficiently as the model indicated. The final modeled illumination of the nadir panel is shown in Figure 16. As was expected, the illumination was larger than the ideal model indicated (Figure 12 above).

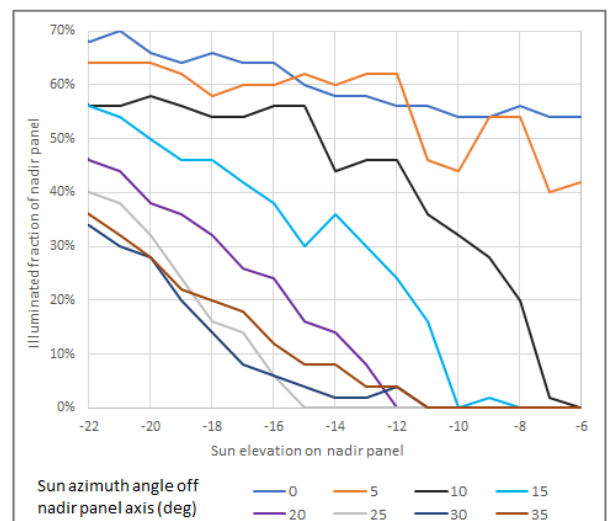

*Figure 16: Fraction of the nadir panel illuminated by the Sun, as fitted by the data. Sun angles off the velocity-vector from 0 to 35° shown individually.*

Final plots - Including the earth-light element, the model predictions and the observed BlueBird magnitudes are presented in Figure 17. A reasonable level of agreement is achieved across eight magnitudes of observed brightness with a root-mean-square error of 0.6ᵐ. The beta-prime angle at the time of each observation is indicated in color-coded bands.

The added scatter at lower magnitudes comes in part from the difficulty of accurately modeling the reflection from the nadir panel when the Sun elevation is near zero. In that condition, the path of light through the large number of stand-offs is particularly complex.

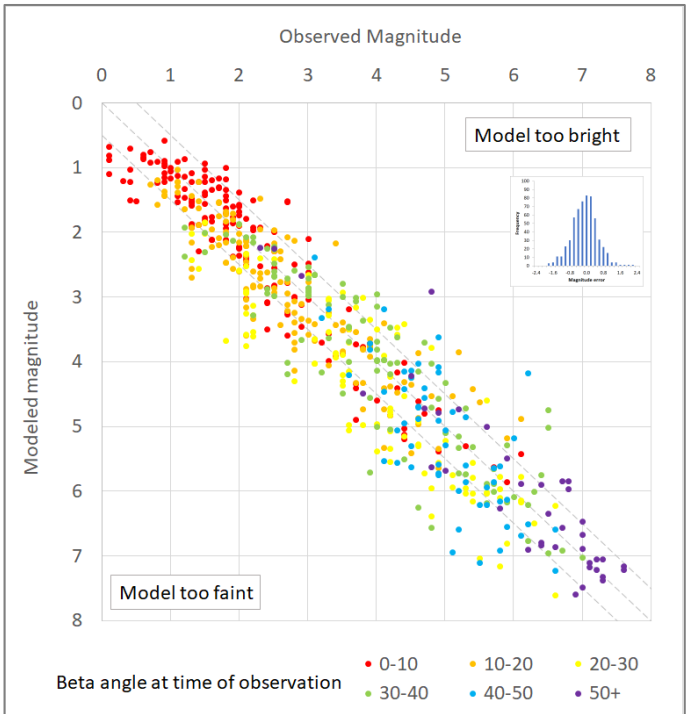

*Figure 17: A comparison of the model predictions against the brightnesses measured in 576 visual observations across the five BlueBird spacecraft, with the final error distribution. The dotted lines indicate ±0.5mag.*

### 5.4  Conclusions from the model

The following conclusions are drawn from the model as fitted to the data.

Zenith Magnitude – Variation of the brightness of the BlueBird spacecraft is dominated by the beta-prime angle at the time of observation. The model can be used to calculate the brightness as a function of the beta angle, elevation and azimuth and the position of the Sun at the observer.

When the beta angle is small, BlueBird is bright and is observed at around 0.5ᵐ in the zenith. When the beta angle is larger, over 30°, BlueBird is fainter in the zenith (Figure 18). At Sun elevation angles ≥-12°, the reflected earth-light element becomes apparent.

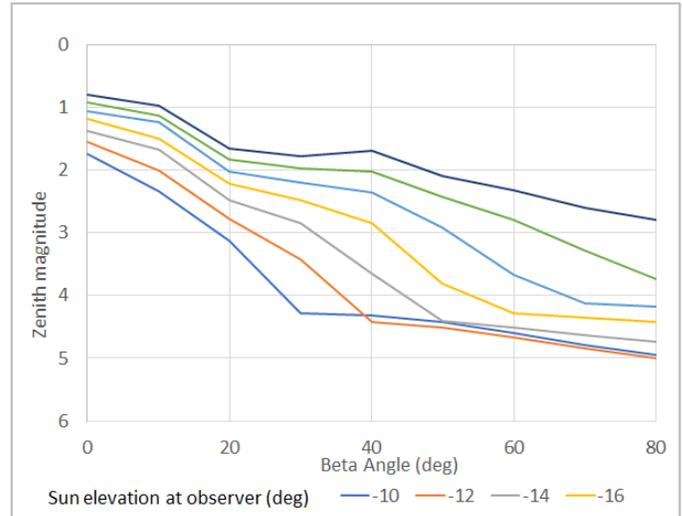

*Figure 18: Zenith magnitudes against the orbit beta angle, as a function of the Sin elevation at the observer*

Brightness across the sky - Figure 19 shows the modeled brightness across the sky for four different simulated BlueBird passes at latitude 40°N. High and low beta angle cases are displayed for two different Sun elevations. In general, the objects are slightly brighter in the anti-Sun direction (low phase angle) than towards the Sun. This is because more reflected light is visible from the sunward side of the stand-offs and the Sun is lower in the sky at the spacecraft, so solar illumination of the nadir panel is greater (less effective shadowing). The high beta angle cases are fainter than the low beta angle cases, particularly when the Sun elevation is larger and the effect of the spacecraft roll on the nadir face brightness is greater. However, the predicted magnitude is never outside easily detectable limits.

### 5.5  Comparison with BlueWalker 3

The analysis in section 3 compared the previous observations of BlueWalker 3 with the new dataset for BlueBird, with the conclusion that there is not much difference in *average* brightness between the two spacecraft.

Given that the stand-off structures provide some shading of the nadir face of the BlueBird spacecraft, some reduction in brightness might be expected, but is not evident in the observed averages reported in section 3.

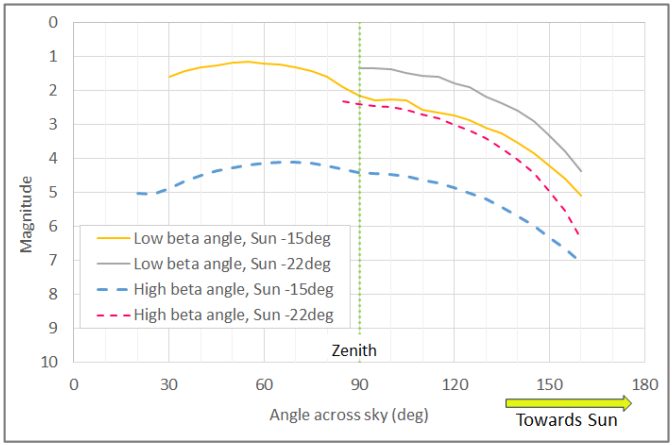

*Figure 19: The modeled BlueBird brightness across the sky from elevation 20° through zenith to elevation 20° in the opposite azimuth, in the pro-Sun direction. High and low beta angle cases are shown for Sun elevations -15° and -20°.*

By suitable choice of parameters, the model can be made to represent the BW3 configuration and results compared to the BlueBird configuration discussed above. It is assumed here that BW3 was tilted using roughly the same roll axis law as BlueBird has been using. While a sunward tilt of BW3 was recognized (Mallama et al, 2023 and 2024), the precise roll law was not identified at that time.

Figure 20 compares the modeled zenith magnitudes of BlueBird and BW3 for three Sun elevations at the observer.

As BW3 rolled around the velocity-vector, the single brightness effect was that angle of incidence of sunlight on the nadir panel increased. When the angles of incidence exceed 90°, the nadir face became dark (apart from earth-light if close to the terminator). It was a feature of BW3 that at high beta angles the spacecraft was dark across a large part of the sky, particularly in bright twilight.

The beta angle changed the azimuth of sunlight falling on the BW3 nadir panel but since that appears to behave as a Lambertian surface, this had no measurable effect on the reflected light.

However, as BlueBird rolls around the velocity-vector there are three significant effects:

1.  The same change of sunlight angle of incidence on the nadir panel as in the BW3 case.

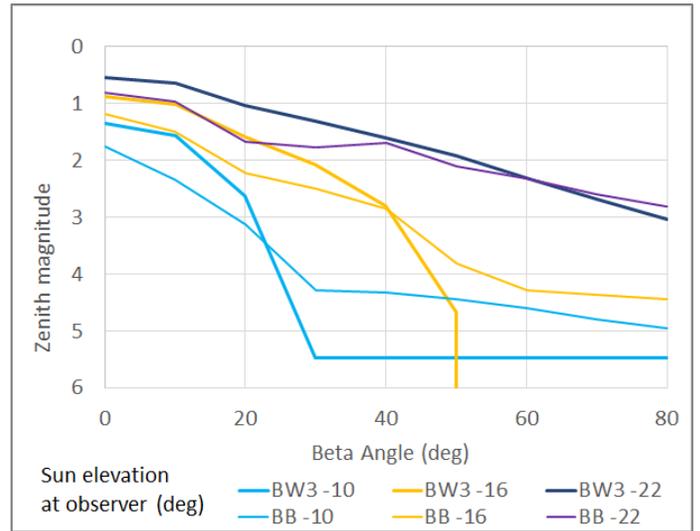

*Figure 20: Comparison of modeled BlueBird and BW3 zenith magnitudes for three Sun elevations at the observer. Thin lines are BlueBird cases, thick lines are BW3.*

2.  The change of sunlight azimuth on the nadir panel increases the shading by the stand-offs.

3.  The roll angle affects the visibility of the side panels (that is the vertical faces of the standoffs) for the observer. In the zenith, the faces of the stand-offs become increasingly visible as the roll angle increases. This acts on the brightness in the opposite sense to effect #2.

Further, the stand-offs are visible at lower elevations in the anti-Sun direction (low phase angle), an increase in brightness that will affect the average.

## 6    Comparison to other spacecraft

This section compares the brightness of BlueBird satellites to other constellations and assesses their impact on observational astronomy.

Table 1 shows that the BlueBird constellation is brightest in terms of apparent and 1000-km magnitudes. The Starlink values are current statistics from our database, while the OneWeb means are from Mallama (2022b).

Satellites brighter than magnitude 7 seriously degrade images from the wide field LSST survey of the Vera Rubin observatory, as described by Tyson et al (2002). For casual sky watchers the limit is magnitude 6 because more luminous satellites are visible to the unaided eye. BlueBird spacecraft are much brighter than these limits.

| Table 1: Comparison of spacecraft magnitudes | | |
|---|---|---|
| Satellite | Apparent magnitude | 1000-km magnitude |
| BlueBirds | 3.44 | 3.84 |
| Starlink-Direct-to-Cell (DTC) | 4.98 | 6.15 |
| Starlink-Mini | 6.36 | 7.22 |
| OneWeb | 7.85 | 7.05 |

## 7    Conclusions

The BlueBird design includes a feature not present in the BlueWalker 3 spacecraft launched in 2022. This addition comprises a set of structures that almost completely shadow the Earth-facing antenna panel when the angle of the Sun with respect to the spacecraft is in a certain range. Under these ideal conditions the zenith brightness of BlueBird is fainter than $4^m$. When the Sun angle conditions are not ideal, BlueBird is of similar zenith magnitude to BlueWalker 3, at brightest $0.5^m$ when the Sun is $22°$ below the horizon just before the spacecraft enters eclipse.

The observations reported here were made in the period September 2024 to March 2025. It is possible that a different attitude control mode will be used in the future for normal BlueBird operations. If so, the Sun angles may lie in the ideal conditions more of the time during normal operations and the peak and average BlueBird brightnesses will be fainter.

The planned BlueBird block 2 spacecraft will be over three times larger than the block 1 units. If a similar sunshade on the block 2 spacecraft is not used efficiently, they will be $1.3^m$ brighter than block 1, close to a visual magnitude of $-1^m$, for a considerable part of the time.